\documentclass[10pt,a4paper,aps,twocolumn,showpacs,amssymb,amsmath, color,floatfix]{revtex4}
\usepackage{ucs}
\usepackage{amsmath}
\usepackage{amsfonts}
\usepackage{amssymb}
\usepackage{graphicx}
\graphicspath{{EPS_figs/}{PDF_figs/}}
\usepackage{color}

\begin{document}
\title{ High order elastic terms, boojums and general paradigm of the elastic interaction between colloidal particles in the nematic liquid crystals.}
\author{S. B. Chernyshuk $^{1)}$}
\affiliation{ $^{1)}$ Institute of Physics, NAS Ukraine, Prospekt
Nauki 46, Kyiv 03650, Ukraine }
\thanks{stasubf@gmail.com}

\begin{abstract}
Theoretical description of the elastic interaction between colloidal particles in NLC with incorporation of the higher order elastic terms beyond the limit of dipole and qudrupole interactions is proposed. The expression for the elastic interaction potential between axially symmetric colloidal particles, taking into account of the high order elastic terms, is obtained. The general paradigm of the elastic interaction between colloidal particles in NLC is proposed so that every particle with strong anchoring and radius $a$ has three zones surrounding itself. The first zone for $a<r\lessapprox 1.3a$ is the zone of topological defects; the second zone at the approximate distance range $1.3a \lessapprox r \lessapprox 4a$ is the zone where crossover from topological defects to the main multipole moment takes place. The  higher order elastic terms are essential nere (from 10\% to 60\% of the total deformation). The third zone is the zone of the main multipole moment, where higher order terms make a contribution of less than 10\%. This zone extends to distances $r\gtrapprox 4a=2D$.

The case of spherical particles with planar anchoring conditions and boojums at the poles is considered as an example.  It is found that boojums can be described analitically via multipole expansion with accuracy up to $1/r^{7}$ and the whole spherical particle can be effectively considered as the multipole of the order 6 with multipolarity equal $2^{6}=64$. The correspondent elastic interaction with higher order elastic terms gives the angle $\theta_{min}=34.5^{\circ}$ of minimum energy between two contact beads which is close to the experimental value of $\theta_{min}=30^{\circ}$. In addition, high order elastic terms make the effective power of the repulsive potential to be non-integer at the range $4.5<\gamma_{eff}<5$ for different distances. The incorporation of the high order elastic terms in the confined NLC produce results that agree with experimental data as well. 

\end{abstract}

\maketitle

\section{Introduction}

Anisotropic properties of the nematic liquid crystals (NLC) give rise to a new class of colloidal elastic anisotropic interactions that never occur in isotropic hosts and result in different structures of colloidal particles: linear chains \cite{po1,po2}, inclined chains with respect to the director  \cite{po3}-\cite{lavr2} and quasi 2D nematic colloids \cite{nych}-\cite{ulyana}. Theoretical understanding of the matter in the bulk NLC is based on the multipole expansion of the director field and has deep electrostatic analogies. Untill now, all theoretical models dealt with only the first three terms in multipole expansion: Coulomb-like \cite{lev3}, dipole and quadrupole \cite{lupe}-\cite{we4}. Almost all experiments are made with axially symmetric colloidal particles (primarily spherical) which carry only dipole and/or quadrupole elastic moments. But considering only these two terms cannot explain quantitatively  any of the observed structures. For instance the droplets with tangential boundary conditions make an angle of $30^{\circ}$ with the alignment axis of the liquid crystal \cite{po3,lavr1,kot}, which is along the vertical axis. However, the quadrupole interaction gives the angle, for which long-range attraction
is maximized to be approximately $49^{\circ}$. Therefore the origin of the existing structures must be ascribed to the short-range effects, not explicitly included in the theory. 

In the current paper it is found that high order multipole terms play a very important role in the short-range effects and in the formation of colloidal structures. Actually we find that there are three zones around each colloidal particle: the first zone is the zone of topological defects where non-linear terms are essential. It has the approximate size of $0.2a-0.3a$ of the particle radius $a$, so that it is concentrated on the distances $a\lessapprox r \lessapprox 1.3a$. The second zone is the intermediate zone, where all possible from the symmetry point of view elastic terms, are born simultaneously and higher order elastic terms are essential (from 10\% to 60\% of the total deformations). It is concentrated at the approximate distances $1.3a \lessapprox r \lessapprox 4a$. And the third zone is the zone of the last multipole moment, where higher order terms make contribution less than 10\% and only the last multipole moment has the dominant value. This third zone extends to distances $r \gtrapprox4a=2D$.

We will now consider the case of spherical particles with planar anchoring conditions as an example. Such particles have topological defects called boojums at the poles. We find that boojums can be effectively described via multipole expansion with accuracy up to $1/r^{7}$ and the whole spherical particle can be effectively considered as multipole of the order 6 with multipolarity equal $2^{6}=64$. The correspondent elastic interaction between two beads with higher order elastic terms gives the angle $\theta_{min}=34.5^{\circ}$ of minimum energy between two contact beads which is close to the experimental value of $\theta_{min}=30^{\circ}$.

\section{Incorporation of the higher order elastic terms into the theory}

Let's now consider axially symmetric particle of the micron or sub-micron size which may carry topological defects such as hyperbolic hedgehog, disclination ring or boojums. In the absence of the particle the non-deformed state of NLC is the orientation of the director $\textbf{n}|| z, \textbf{n}=(0,0,1)$. The immersed particle induces deformations of the director in the perpendicular directions $n_{\mu}, \mu=x,y$ and make director field $\textbf{n}\approx(n_{x},n_{y},1)$. The bulk energy of deformation may be approximately written in the harmonic form:
\begin{equation}
F_{har}=\frac{K}{2}\int  d^{3}x (\nabla n_{\mu})^{2}\label{fb3}
\end{equation}
with Euler-Lagrange equations of Laplace type:
\begin{equation}
\Delta n_{\mu}=0  \label{lap}
\end{equation}
Then the director field outside the particle in the infinite LC has the form  $n_{x}(\textbf{r})=p\frac{x}{r^{3}}+3c\frac{xz}{r^{5}},n_{y}(\textbf{r})=p\frac{y}{r^{3}}+3c\frac{yz}{r^{5}}$ in the simplest case with $p$ and $c$ being dipole and quadrupole elastic moments. The anharmonic correction to the bulk energy is $F_{anhar}=\frac{K}{2}\int  d^{3}x (\nabla n_{z})^{2}\approx\frac{K}{8}\int  d^{3}x (\nabla n_{\bot}^{2})^{2}$ which changes EL equations to be:
\begin{equation}
\Delta n_{\mu}+\frac{1}{2}n_{\mu}\Delta n_{\bot}^{2}=0  \label{unhar}
\end{equation}
If the leading contribution to $n_{\mu}$ is the dipolar term then anharmoic corrections are of the form $r_{\mu}/r^{7}$ and high order terms of the order up to $1/r^{5}$ can effectively influence on the short-range behaviour and should be equally considered. The same if the leading contribution to $n_{\mu}$ is the quadrupolar term then anharmonic corrections are of the form $r_{\mu}/r^{10}$ and high order terms of the order up to $1/r^{8}$ can effectively influence the short-range behaviour.

In the general case, the solution of the Laplace equation for axially symmetric particles has the form: 
\begin{equation}
n_{\mu}=\sum^{N}_{l=1}a_{l}(-1)^{l}\partial_{\mu}\partial_{z}^{l-1}\frac{1}{r}  \label{hn}
\end{equation}
where $a_{l}$ is the multipole moment of the order $l$ and $2^{l}$ is the multipolarity; $N$ - is the maximum possible order without anharmonic corrections. For the dipole particle $N=4$, for the quadrupole particle $N=7$. So $a_{1}=p$ is the dipole moment, $a_{2}=c$ - is the quadrupole moment. Actually, all odd coefficients are equal to zero for quadrupole particles $a_{3}=a_{5}=a_{7}=0$  because of the horisontal symmetry plane so that it can be limited with $N=6$. All nonzero coefficiants $a_{l}$ are unknown quantities. They can be found as asymptotics from exact solutions or from variational ansatzes. Strictly speaking these coefficients are functions of the anchoring coefficient $W$ and surface elastic constants $a_{l}=a_{l}(Wa/K,K_{24}/K,K_{13}/K)$ as well as the Frank elastic constant $K$ and particle radius $a$. But we don't exactly know this dependence. From the other side the coefficients $a_{l}$ may be found as fitting parameters in the interaction potentials for each particular case. 

 In order to find the energy of the system: particle(s) + LC , it is necessary to introduce some effective free energy functional $F_{eff}$ so that it's Euler-Lagrange equations would have the above solutions (\ref{hn}). In the one constant approximation with Frank constant $K$ the effective functional has the form:
\begin{equation}
F_{eff}=K\int d^{3}x\left\{\frac{(\nabla n_{\mu})^{2}}{2}-4\pi\sum^{N}_{l=1}A_{l}(\textbf{x})\partial_{\mu}\partial_{z}^{l-1}n_{\mu} \right\}\label{flin}
\end{equation}
which brings Euler-Lagrange equations:
\begin{equation}
\Delta n_{\mu}=4\pi\sum^{N}_{l=1}(-1)^{l-1}\partial_{\mu}\partial_{z}^{l-1}A_{l}(\textbf{x})\label{nmu}
\end{equation}
where $A_{l}(\textbf{x})$ are multipole moment densities, $\mu=x,y$ and repeated $\mu$ means summation on $x$ and $y$ like $\partial_{\mu}n_{\mu}=\partial_{x}n_{x}+\partial_{y}n_{y}$.
For the infinite space the solution has the known form: 
\begin{equation}
n_{\mu}(\textbf{x})=\int d^{3}\textbf{x}' \frac{1}{\left|\textbf{x}-\textbf{x}'\right|}\sum^{N}_{l=1}(-1)^{l}\partial_{\mu}'\partial_{z}'^{l-1}A_{l}(\textbf{x}') \label{solmain} 
\end{equation}

\begin{figure}
\begin{center}
\includegraphics[width=\columnwidth]{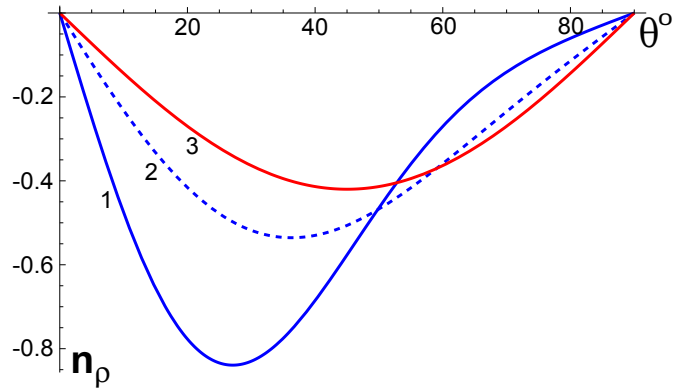}
\caption{(Color online) Horizontal projection of the director $n_{\rho}=n_{\bot}$ at the surface of the spherical particle $r=a$ with planar anchoring conditions for $0\leq\theta\leq 90^{\circ}$ (see Fig.\ref{gdefects} d). According to the solution (\ref{nx}). Blue line 1 corresponds to $(b_{2},b_{4},b_{6})=(-0.36, -0.023, -0.00018)$, blue dashed line 2 corresponds to $b_{6}=0$, $(b_{2},b_{4},b_{6})=(-0.32, -0.007, 0)$ and red line 3 corresponds to the pure quadrupole term  for $b_{4},b_{6}=0$, $(b_{2},b_{4},b_{6})=(-0.28,0,0)$.\label{gnro}}
\end{center}
\end{figure}

\begin{figure}
\begin{center}
\includegraphics[width=\columnwidth]{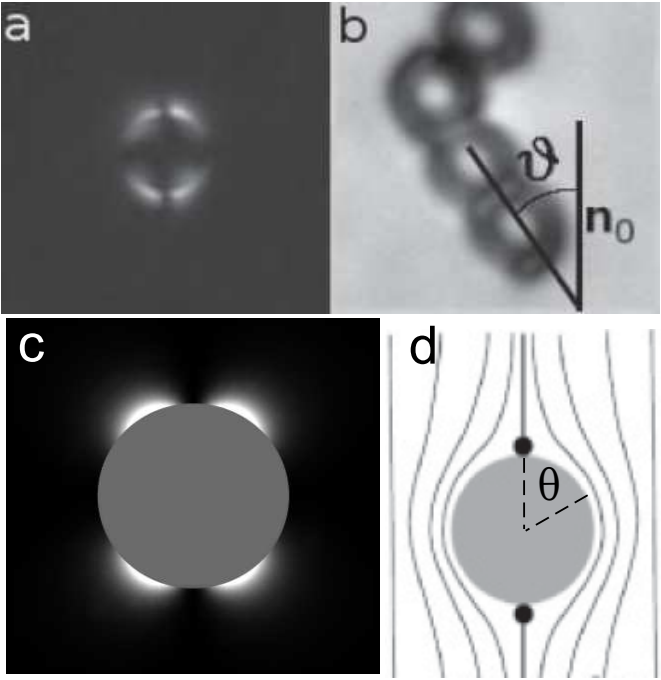}
\caption{(a) Taken from \cite{kot}. Colloidal particle ($2a=4.5 \mu m$) immersed in
nematic liquid crystal in a planar cell as seen under a polarizing microscope. Two boojums at the poles of the sphere confirm tangential alignment of the director at the sphere surface. (b) Taken from \cite{kot}. Aggregation of particles was observed at an angle of approximately $\theta\approx 25^{\circ}-35^{\circ}$ with respect to the average director orientation $\textbf{n}_{0}||z$.(c) Intensity profile of the nematic field obtained by numerical calculation shown by grayscale plots of $n_{\rho}^{2}$ from the solution (\ref{nx}) for $(b_{2},b_{4},b_{6})=(-0.36, -0.023, -0.00018)$. In the black regions the director aligns along the z axis, while deformations $n_{\rho}^{2}$ are maximal in the white regions. (d) Chart of the director field for the boojums configuration.\label{gdefects} }
\end{center}
\end{figure}

If we consider $A_{l}(\textbf{x})=a_{l}\delta(\textbf{x})$ this really brings solution (\ref{hn}). This means that effective functional (\ref{flin}) correctly describes the interaction between the particle and LC. 

 Consider $N_{p}$ particles in the NLC, so that $A_{l}(\textbf{x})=\sum_{i}a_{l}^{i}\delta(\textbf{x}-\textbf{x}_{i})$, $i=1\div N_{p}$ . Then substitution (\ref{solmain}) into $F_{eff}$ (\ref{flin}) brings: $F_{eff}=U^{self}+U^{interaction}$ where $U^{self}=\sum_{i}U_{i}^{self} $ , here $U_{i}^{self}$ is the divergent self energy.\\
Interaction energy $ U^{interaction}=\sum_{i<j}U_{ij}^{int} $.  Here $U_{ij}^{int}$ is the elastic interaction energy between $i$ and $j$ particles in the unlimited NLC: 
\begin{equation}
U_{ij}^{int}=4\pi K \sum^{N}_{l,l'=1}a_{l}a_{l'}'(-1)^{l'}\frac{(l+l')!}{r^{l+l'+1}}P_{l+l'}(cos\theta)\label{uint}
\end{equation}
Here unprimed quantities $a_{l}$ are used for particle $i$ and primed $a_{l'}'$ for particle $j$, $r=|\textbf{x}_{i}-\textbf{x}_{j}|$, $\theta$ is the angle between $\textbf{r}$ and z and we used the relation $P_{l}(cos\theta)=(-1)^{l}\frac{r^{l+1}}{l!}\partial_{z}^{l}\frac{1}{r}$ for Legendre polynomials $P_{l}$. It is the general expression for the elastic interaction potential between axially symmetric colloidal particles in the unbounding NLC with taking into account of the high order elastic terms.

 The case of confined NLC means just replacement of $\frac{1}{r}=\frac{1}{|\textbf{x}_{i}-\textbf{x}_{j}|}$ with the Green's function $G(\textbf{x},\textbf{x}')$ (see \cite{we,we2}) , which satisfies equation $\Delta_{\textbf{x}}G(\textbf{x},\textbf{x}')=-4\pi \delta(\textbf{x}-\textbf{x}')$ for  $\textbf{x},\textbf{x}'\in \textbf{V}$ ($\textbf{V}$ is the volume of the bulk NLC) and $G(\textbf{x},\textbf{s})=0 $ for any $\textbf{s}$ of the bounding surfaces $\Sigma$.
Then formula (\ref{uint}) for the confined NLC has the form:
\begin{equation}
U_{ij}^{int,confined}=-4\pi K\sum^{N}_{l,l'=1}a_{l}a_{l'}'\partial_{\mu}\partial_{\mu}'\partial_{z}^{l-1}\partial_{z}'^{l'-1}G(\textbf{x}_{i},\textbf{x}_{j}')\label{uintconf}
\end{equation}
For dipole particles the sum is limited to nonzero $a_{1},a_{2},a_{3}$ and $a_{4}$. For quadrupole particles (beads with boojums and Saturn ring configuration) the sum is limited to nonzero $a_{2},a_{4}$ and $a_{6}$. All coefficients may be presented as $a_{l}=b_{l}a^{l+1}$ with $a$ being the radius of the particle and $b_{l}$ are just dimensionless parameters.

Below we consider a spherical particle with planar anchoring conditions at the surface as an example. Then the director field (\ref{hn}) can be presented as $n_{\mu}=a_{2}\partial_{\mu}\partial_{z}\frac{1}{r}+a_{4}\partial_{\mu}\partial_{z}^{3}\frac{1}{r}+a_{6}\partial_{\mu}\partial_{z}^{5}\frac{1}{r}$. Let's introduce dimensionless distance $r\Rightarrow a r$, then the horizontal projection $n_{\rho}=n_{\bot}$ ($n_{\rho}^{2}=n_{\mu}n_{\mu}$) of the director has the form:

\begin{equation}\label{nx}
\begin{gathered}
n_{\rho}=b_{2}\frac{3sin\theta cos\theta}{r^{3}}+b_{4}\frac{105sin\theta cos^{3}\theta-45sin\theta cos\theta}{r^{5}}+\\
+b_{6}\frac{10395sin\theta cos^{5}\theta - 9450 sin\theta cos^{3}\theta + 1575 sin\theta cos\theta}{r^{7}}
\end{gathered}
\end{equation}
where $\theta$ is the angle between $z$ and r, $n_{x}=n_{\rho}cos\varphi$, $n_{y}=n_{\rho}sin\varphi$ and $\varphi$ is the angle between $\rho$ and $x$. It is obvious that $b_{2}<0$ for the boojums configuration.  In the paper \cite{we2} it was found that $b_{2}\approx -0.28$ for the experiment \cite{conf} and we have two unknown variation parameters $b_{4}$ and $b_{6}$. Physical limitations for these coefficients may be formulated in the following way: $|n_{\rho}|<1$ for all distances $r\geq1$; $n_{\rho}(r,\theta)$ should have only one minimum/maximum as a function of $\theta$ for $0<\theta<\pi/2$ and for all $r\geq1$ and the correspondent energy of interaction (\ref{uint}) should agree with all known experimental results as much as possible. We found that all these conditions are satisfied in the best way for the values $(b_{2},b_{4},b_{6})=(-0.36,-0.023,-0.00018)$(see below). 

The dependence $n_{\rho}(\theta)$ on the spherical surface $r=1$ is depicted on the Fig.\ref{gnro} for the values $(b_{2},b_{4},b_{6})=(-0.36,-0.021,-0.00011)$ (blue line 1). Topological defects called boojums are located at the poles of the particle (see Fig.\ref{gdefects}.a). The intensity profile of the nematic field obtained by numerical calculation shown by grayscale plots of $n_{\rho}^{2}$ from the solution (\ref{nx}) is presented on the Fig.\ref{gdefects}.c . In the black regions the director aligns along the z axis, while deformations $n_{\rho}^{2}$ are maximal in the white regions. We see that pictures  Fig.\ref{gdefects}.a and Fig.\ref{gdefects}.c are quite similar. This means that solution (\ref{nx}) gives the correct analitical description of the boojums near the surface of the spherical particles up to the $1/r^{7}$ order. Of course there will be some corrections to the solution from the anharmonic term, but they will decrease faster than $1/r^{7}$ as it is seen from the equation (\ref{unhar}). 

We will discuss, in more detail, the limits of applicability of the higher order elastic terms as well as make a more profound estimation of nonlinear terms in the Sec III.
\begin{figure}
\begin{center}
\includegraphics[width=\columnwidth]{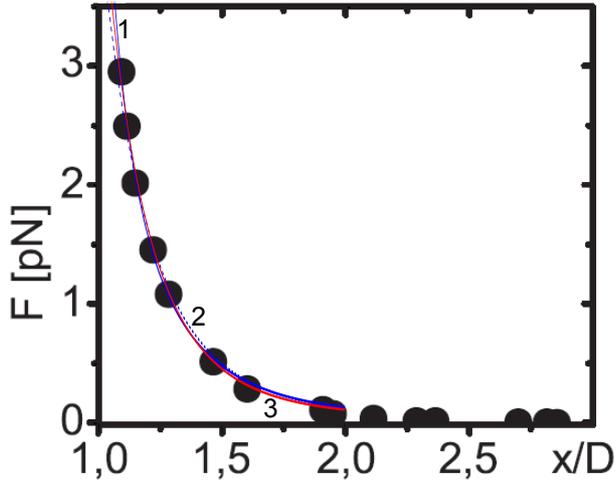}
\caption{(Color online) Experimental values (taken from \cite{conf}) of interparticle repulsion force F between two beads of diameter $D=2a=4.4 \mu m$ as a
function of rescaled distance $x/D$. Blue line 1 is calculated as $F=-U'_{r}$ from (\ref{uboojum}) for  $\theta=\pi/2$, $K=7 pN$ (5CB) and  $(b_{2},b_{4},b_{6})=(-0.36, -0.023, -0.00018)$, blue dashed line 2 corresponds to $b_{6}=0$, $(b_{2},b_{4},b_{6})=(-0.32, -0.007, 0)$ and red line 3 corresponds to the pure quadrupole-quadrupole interaction for $b_{4},b_{6}=0$, $(b_{2},b_{4},b_{6})=(-0.28,0,0)$ .\label{force}}
\end{center}
\end{figure}

\begin{figure}
\begin{center}
\includegraphics[width=\columnwidth]{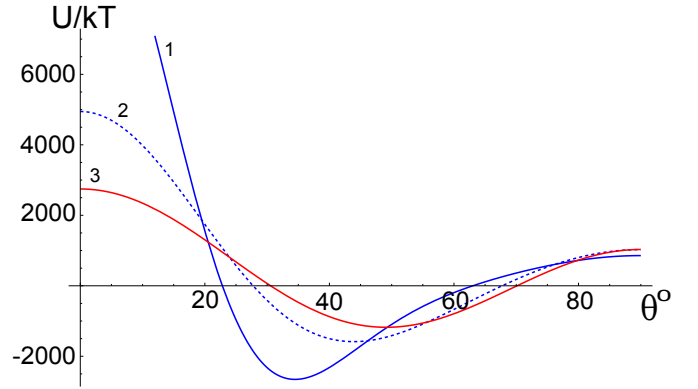}
\caption{(Color online) Blue line 1 corresponds to the angular dependence of the interaction potential (\ref{uboojum}) in kT units for two close beads at the distance $r=2a$, $a=2.2 \mu m$, $K=7 pN$ for $(b_{2},b_{4},b_{6})=(-0.36, -0.023, -0.00018)$. The potential has minimum at $\theta=34.5^{\circ}$. Blue dashed line 2 corresponds to the case $b_{6}=0$, $(b_{2},b_{4},b_{6})=(-0.32, -0.007, 0)$. The potential has minimum at $\theta=44^{\circ}$.  Red line 3 corresponds to the pure quadrupole-quadrupole interaction for $b_{4},b_{6}=0$, $(b_{2},b_{4},b_{6})=(-0.28,0,0)$ with minimum at $\theta=49^{\circ}$ .\label{guang}}
\end{center}
\end{figure}

\begin{figure}
\begin{center}
\includegraphics[width=\columnwidth]{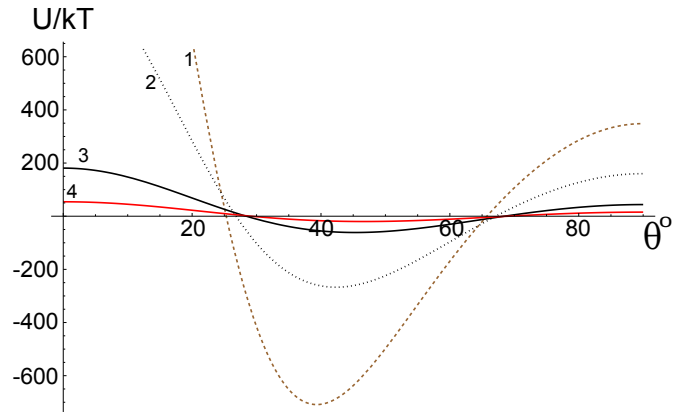}
\caption{(Color online) The angular dependence of the interaction potential (\ref{uboojum}) in kT units for two  beads at different distances. Broun dashed line 1 corresponds to  $r=2.5a$, black dotted line 2 for $r=3a$, black line 3 for $r=4a$, red line 4 for $r=5a$. Here radius $a=2.2 \mu m$, $K=7 pN$ and  $(b_{2},b_{4},b_{6})=(-0.36,-0.023,-0.00018)$.\label{angpot}}
\end{center}
\end{figure}

\begin{figure}
\begin{center}
\includegraphics[width=\columnwidth]{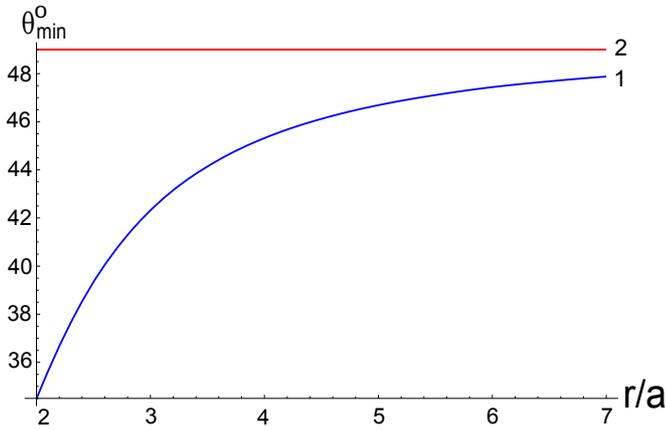}
\caption{(Color online) Dependence of the minimum energy angle $\theta_{min}$ on the rescaled distance $r/a$ between two beads for the potential (\ref{uboojum}) with $(b_{2},b_{4},b_{6})=(-0.36,-0.023,-0.00018)$ (blue line 1), minimum energy angle $\theta=49^{\circ}$ for pure quadrupole interaction (red line 2).\label{angmin}}
\end{center}
\end{figure}

\subsection{The effect of the high order terms on the angular dependence of the interaction potential}

The correspondent energy (\ref{uint}) of elastic interaction between two spheres with boojums has the form (all coefficients $b_{l}=0$ besides  $(b_{2},b_{4},b_{6})$ and $N=6$):
\begin{equation}\label{uboojum}
\begin{gathered}
\frac{U}{4\pi aK}= b_{2}^{2}4!\frac{P_{4}(cos\theta)}{r^{5}}+2b_{2}b_{4}6!\frac{P_{6}(cos\theta)}{r^{7}}\\
+(2b_{2}b_{6}+b_{4}^{2})8!\frac{P_{8}(cos\theta)}{r^{9}}+2b_{4}b_{6}10!\frac{P_{10}(cos\theta)}{r^{11}}+\\
+b_{6}^{2}12!\frac{P_{12}(cos\theta)}{r^{13}}
\end{gathered}
\end{equation}

where $r$ is measured in radius $a$ units and $r\geq2$. The correspondent force $F=-U'_{r}$ of repulsion between two beads of diameter $D=2a=4.4 \mu m$ in 5CB ($K=7pN$) for $\theta=\pi/2$ is plotted on the Fig.\ref{force}. We see that experimental values of the repulsion force may be fitted with three different sets of parameters: blue line 1 corresponds to  $(b_{2},b_{4},b_{6})=(-0.36, -0.023, -0.00018)$; blue dashed line 2 corresponds to the case $b_{6}=0$,  $(b_{2},b_{4},b_{6})=(-0.32, -0.007, 0)$. Red line 3 corresponds to the pure quadrupole-quadrupole interaction for $b_{4},b_{6}=0$, $(b_{2},b_{4},b_{6})=(-0.28,0,0)$. It is very interesting that all these three set of parameters fit the data very well on the distances $1<r/D<1.6$. Actually the experimental values on the Fig.\ref{force} were found in the homeotropic cell with width $L=1.8D=7.9 \mu m$ \cite{conf}. It was found in \cite{conf,fukuda,we} that confining effects become essential for distances of more than $r>0.9L$ so that we can use approximation of the unbounding NLC (\ref{uboojum}) for the distances $r<0.9L=1.6D$ (for $L=1.8D$).

Despite the fact that three different set of parameters give almost the same values of the repulsion force in the perpendicular direction  $\theta=\pi/2$, they produce \textit{very different pictures} for the angular dependence of the interaction potential. The angular dependences of the interaction potential for two close contact beads at the distance $r=2$ is depicted on the Fig.\ref{guang}. It is clearly seen that two beads with set of parameters $(b_{2},b_{4},b_{6})=(-0.36, -0.023, -0.00018)$ (Fig.\ref{guang}, blue line 1) produce the interaction potential (\ref{uboojum}) which has the minimum at the angle $\theta=34.5^{\circ}$ that is very close to the results observed earlier in experiments \cite{po3,lavr1} (see Fig.\ref{gdefects}.b). The set of parameters  $(b_{2},b_{4},b_{6})=(-0.32, -0.007, 0)$ ( Fig.\ref{guang}, blue dashed line 2) produces the potential with the minimum at the angle $\theta=44^{\circ}$. Red line 3 corresponds to the pure quadrupole-quadrupole interaction for $b_{4},b_{6}=0$, $(b_{2},b_{4},b_{6})=(-0.28,0,0)$ and the correspondent interaction potential (\ref{uboojum}) has minimum at $\theta=49^{\circ}$. 

The angular dependence of the elastic interaction potential for different distances $r$ is depicted on the Fig.\ref{angpot}. The minimum energy angle increases from $\theta=34.5^{\circ}$ to $\theta=48^{\circ}$ with increase of the distance from $r=2a$ to $r=8a$ that is shown on the Fig.\ref{angmin}. This corresponds to the experimental results of \cite{lavr1} where minimum energy angle $\theta$ was found to be changed from $\theta=30^{\circ}$ to $\theta=48^{\circ}$ with increase of the distance between particles. 

So we come to the conclusion that high order elastic terms have very profound influence on the angular dependence of the interaction potentials at the short distances between particles which agree with experimental results.

\subsection{The effect of the high order terms on the effective power}
Simple electrostatic analogy developed in \cite{lupe} predicts that elastic forces are proportional to $F\propto r^{-4},r^{-5},r^{-6}$ for different types of elastic interactions. But many experiments give non-enteger power dependence $F\propto r^{-\delta}$ with $\delta=3.6$ in \cite{jap,jap3}, $\delta=4.6$ in \cite{jap2,jap3} for different director configurations. In the paper \cite{jap3} this descrepancy was succesfully fitted with help of possible contribution of higher order terms in multipole expansion of $F$. We argue, as well, that high order elastic terms make an effective power to be non-integer in the range of severel percents.

Let's consider the potential (\ref{uboojum}) for $\theta=\pi/2$ and the set of parameters $(b_{2},b_{4},b_{6})=(-0.36, -0.023, -0.00018)$. This potential is repulsive elsewhere. Let us present it, approximately, in the form of power law dependence with some effective power that depends on the distance, i.e.:

\begin{equation}
U_{same,hom}^{wall}\approx \frac{C}{r^{\gamma_{eff}}}\label{gamma}
\end{equation}

where $\gamma_{eff}$ may be found as $\gamma_{eff}=-\frac{\partial log U}{\partial log r}=-U'_{r}\frac{r}{U}$. Fig.\ref{gamma} shows dependence of such effective power on the dimensionless distance $r/a$. We see that on small distances $2<r/a<3$ effective power $\gamma_{eff}$ decreases from $5.2$ to $4.5$ and then it increases from $4.5$ to $5$ for $r/a>3$.

\begin{figure}
\begin{center}
\includegraphics[width=\columnwidth]{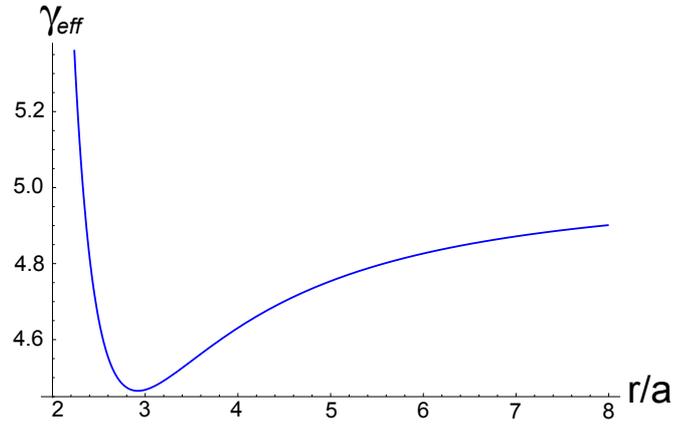}
\caption{(Color online) Dependence of the effective power on the rescaled distance $r/a$ between two beads in the perpendicular direction $\theta=\pi/2$ for the potential (\ref{uboojum}) with $(b_{2},b_{4},b_{6})=(-0.36,-0.023,-0.00018)$.\label{gamma}}
\end{center}
\end{figure}

\subsection{The effect of the high order terms in the confined cell}

The formula (\ref{uintconf}) gives the elastic interaction potential in the confined NLC. Lets's consider homeotropic nematic cell with thickness $L$. The Green function $G_{hom}^{cell}(\textbf{x},\textbf{x}')$ coincides with the Green function $G(\textbf{x},\textbf{x}')$ of the two conducting walls in the electrostatics (see \cite{jac}):

\begin{multline}\label{G_hz}
G_{hom}^{cell}(\textbf{x},\textbf{x}^{\prime})=\frac{4}{L}\sum_{n=1}^{\infty}\sum_{m=-\infty}^{\infty}e^{i m (\varphi-\varphi^{\prime})}\sin\frac{n \pi z}{L}\times\\
\times\sin\frac{n \pi z^{\prime}}{L} I_{m}(\frac{n\pi \rho_{<}}{L})K_{m}(\frac{n\pi \rho_{<}}{L})
\end{multline}

Here heights $z,z'$, horizontal projections $\rho_{<},\rho_{>}$ and $I_{m},K_{m}$ are modified Bessel functions.
Then using of (\ref{uintconf}) brings the elastic interaction between two  beads with boojums in the homeotropic cell :

\begin{multline}\label{uhom}
\frac{U_{cell}^{hom}}{16\pi a K}=\left(\frac{ a}{L}\right)^{5}\sum_{n=1}^{\infty}(n\pi)^{4}cos\frac{n\pi z}{L}cos\frac{n\pi z'}{L}K_{0}(\frac{n\pi \rho}{L})\times\\
 \times [b_{2}^{2}+2b_{2}b_{4}(n\pi)^{2}\left(\frac{a}{L}\right)^{2}+(b_{4}^{2}+2b_{2}b_{6})(n\pi)^{4}\left(\frac{ a}{L}\right)^{4}+\\
+2b_{4}b_{6}(n\pi)^{6}\left(\frac{ a}{L}\right)^{6}+b_{6}^{2}(n\pi)^{8}\left(\frac{ a}{L}\right)^{8} ]
\end{multline}

with $\rho$ being the horizontal projection of the distance between the particles. 

\begin{figure}
\begin{center}
\includegraphics[width=\columnwidth]{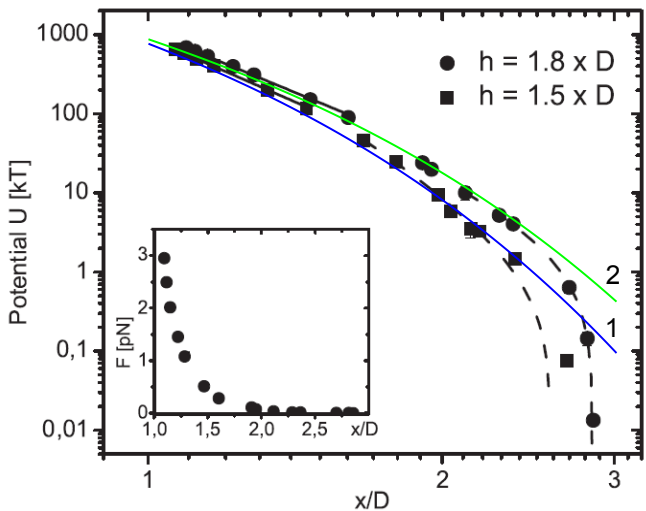}
\caption{(Color online) (Color online)  Experimental data taken from \cite{conf} - energy of elastic interaction between two spherical particles with diameter $D=2a=4.4\mu m$ in the homeotropic cell with thicknesses $L=h=6.5\mu m$ and $h=8\mu m$. The solid blue line 1 and green line 2 are calculated according to the formula (\ref{uhom}) for $z=z'=L/2$. The parameters $(b_{2},b_{4},b_{6})=(-0.36,-0.023,-0.00018)$ fit both thicknesses pretty well in the energy scale $1\textendash1000 kT$.\label{conf}}
\end{center}
\end{figure}

Fig.\ref{conf} demonstrates the application of this formula (\ref{uhom}) for the repulsion potential between two spherical particles (with planar anchoring on the surface providing quadrupole director configuration) with diameter $D=2a=4.4\mu m$ in the center of homeotropic cell ($z=z'=L/2$) with thicknesses $L=6.5\mu m$ and $L=8\mu m$ ( experimental data are taken from \cite{conf} ). It is seen that the set of parameters $(b_{2},b_{4},b_{6})=(-0.36,-0.023,-0.00018)$ fit both thiknesses pretty well in the energy scale $1\textendash1000 kT$.

\section{ The influence of Nonlinear terms. The limits of applicability of the higher order elastic terms. }

In this section we want to discuss the limits of applicability of the higher order elastic terms. In order to do this we need to estimate anharmonic energy term $F_{anhar}=\frac{K}{2}\int  d^{3}x (\nabla n_{z})^{2}=\frac{K}{8}\int  d^{3}x \frac{(\nabla n_{\bot})^{2}}{(1-n_{\bot}^{2})}=\int  d^{3}x f_{anhar}(\textbf{x})$ and compare it with the harmonic term $F_{har}=\frac{K}{2}\int  d^{3}x (\nabla n_{\mu})^{2}=\int  d^{3}x f_{har}(\textbf{x})$. In addition, we need to compare the contribution of the high order elastic terms to the total deformation and compare it with the contribution of the main multipole term.

If we substitute the solution (\ref{nx}) with $(b_{2},b_{4},b_{6})=(-0.36,-0.023,-0.00018)$ into $F_{anhar}$ and $F_{har}$ we receive 
after numerical integration $F_{har}=3.239Ka$, $F_{anhar}=1.065 Ka$ so that total deformation energy is $F_{deform}=4.3Ka$. Let's analize where the anharmonic energy term is the most localized. To do this we introduce the ratio of the anharmonic and harmonic energy densities:

\begin{multline}\label{epsilonf}
\varepsilon_{anharm}(\textbf{x})=\frac{f_{anhar}(\textbf{x})}{f_{har}(\textbf{x})}=\frac{(\nabla n_{\bot}^{2})^{2}}{4(\nabla n_{\mu})^{2}(1-n_{\bot}^{2})}
\end{multline}

\begin{figure}
\begin{center}
\includegraphics[width=\columnwidth]{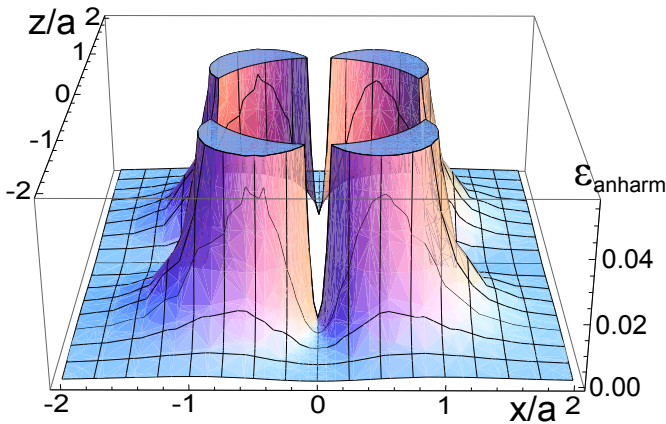}
\caption{(Color online) The ratio of the anharmonic and harmonic energy densities $\varepsilon_{anharm}(\textbf{x})$ (\ref{epsilonf}) in ZX plane around the spherical particle of the unit radius.\label{anharm}}
\end{center}
\end{figure}

\begin{figure}
\begin{center}
\includegraphics[width=\columnwidth]{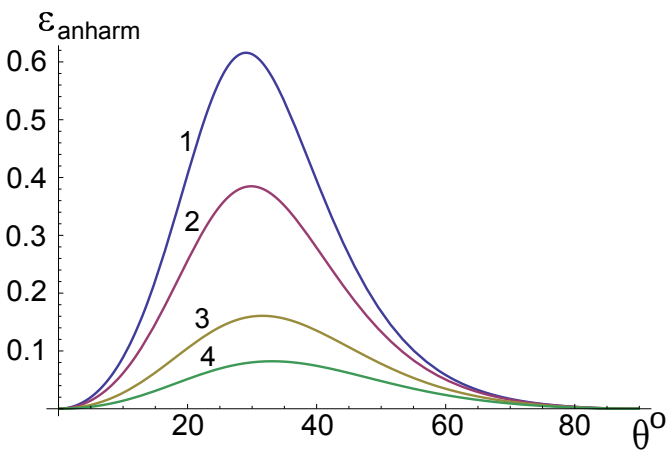}
\caption{(Color online)The ratio of the anharmonic and harmonic energy densities $\varepsilon_{anharm}(\theta,r)$ in the first quarter of the Fig.\ref{anharm} for different distances from the center of the spherical particle ($\theta$ is the angle between $r$ and $z$). Blue line 1 correspond to the distance - $r/a=1.06$, lilac line 2 - $r/a=1.1$, brown line 3 - $r/a=1.2$, green line 4 - $r/a=1.3$  \label{energy_epsilon}}
\end{center}
\end{figure}

 The Fig.\ref{anharm} demonstrates $\varepsilon_{anharm}(z,x)$ in the plane ZX ($ \textbf{n}_{0}||z $). It is clearly seen that $\varepsilon_{anharm}(z,x)$ is localized in the area of $0.2a-0.3a$ from the particle surface and it becomes rapidly less than $10\%$ further (see Fig.\ref{energy_epsilon} as well. )

\begin{figure}
\begin{center}
\includegraphics[width=\columnwidth]{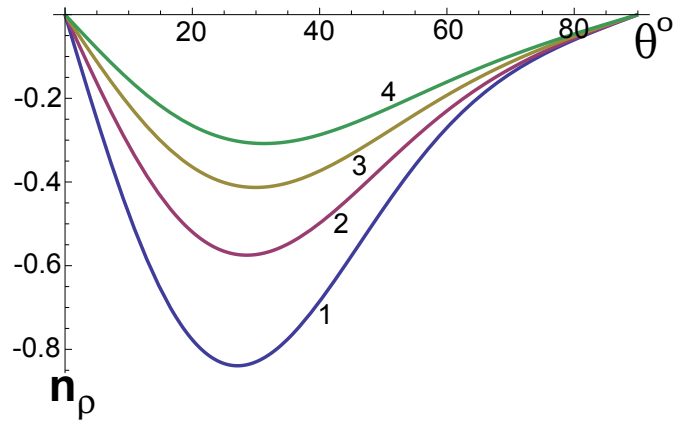}
\caption{(Color online)  Horizontal projection of the director $n_{\rho}=n_{\bot}$ at the different distances from the center of the particle. According to the solution (\ref{nx}) with $(b_{2},b_{4},b_{6})=(-0.36,-0.023,-0.00018)$. Blue line 1 correspond to the distance - $r/a=1$, lilac line 2 - $r/a=1.1$, brown line 3 - $r/a=1.2$, green line 4 - $r/a=1.3$ \label{nrodiff}}
\end{center}
\end{figure}

Fig.\ref{nrodiff} demonstrates perpendicular projection of the director field $n_{\rho}=n_{\bot}$ at the different distances from the center of the particle according to the solution (\ref{nx}) with $(b_{2},b_{4},b_{6})=(-0.36,-0.023,-0.00018)$. It is large enough $0.4<n_{\rho}<0.8$ on the distances $1<r/a<1.2$  and non-linear corrections to the solution (\ref{nx}) may be essential here. But further it becomes smaller $n_{\rho}<0.3$ for $1.3<r/a$ so that condition of harmonic approximation $n_{\mu}\ll 1$ is satisfied and the director is very well described by the solution (\ref{nx}). Thus high order terms are born and are  essential for distances $1.3<r/a$ on the left side. Let's estimate where the end of their influence is on the right side.

Let $n_{\rho}^{h.o.}$ be the part of the director deformation produces by the high order elastic terms (see (\ref{nx})):

\begin{equation}\label{nxh}
\begin{gathered}
n_{\rho}^{h.o.}=b_{4}\frac{105sin\theta cos^{3}\theta-45sin\theta cos\theta}{r^{5}}+\\
+b_{6}\frac{10395sin\theta cos^{5}\theta - 9450 sin\theta cos^{3}\theta + 1575 sin\theta cos\theta}{r^{7}}
\end{gathered}
\end{equation}

Let's introduce the fraction $\eta_{h.o.}$ of the deformation produced by the high order elastic terms in the total deformation:

\begin{equation}\label{eta}
\eta_{h.o.}(\textbf{x})=\left|\frac{n_{\rho}^{h.o.}(\textbf{x})}{n_{\rho}(\textbf{x})}\right|
\end{equation}
where the total deformation $n_{\rho}(\textbf{x})$ is defined in (\ref{nx}).

\begin{figure}
\begin{center}
\includegraphics[width=\columnwidth]{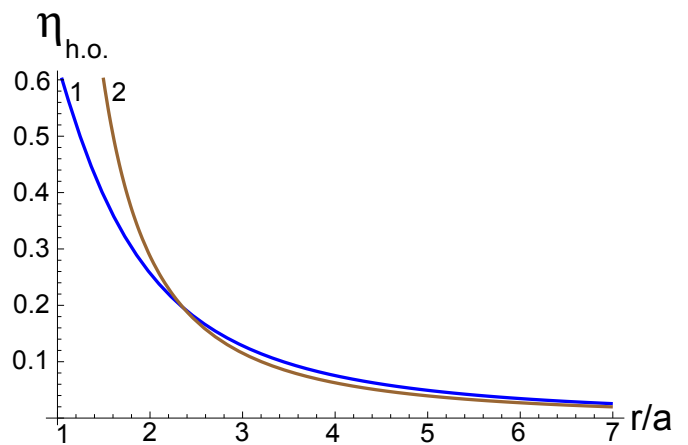}
\caption{(Color online) The fraction of the high order elastic deformations $\eta_{h.o.}(r)$ (\ref{eta}) in the total deformations around the spherical colloidal particle. Blue line 1 corresponds to the direction $\theta=0$, brown line 2 corresponds to the direction $\theta=\pi/2$ \label{ratio}}
\end{center}
\end{figure}

Fig.\ref{ratio} demonstrates the fraction $\eta_{h.o.}(r)$  for two different directions: $\theta=0$ (blue line 1) and $\theta=\pi/2$ (brown line 2) with $(b_{2},b_{4},b_{6})=(-0.36,-0.023,-0.00018)$. It is clearly seen that this fraction is large enough (from $10\%$ to $60\%$) in the range $r/a<4$ and it is less than $10\%$ for $r/a>4$ where the main multipole term (quadrupole in this case) plays the dominant role. Therefore we can say that high order elastic terms play an important role in the range $1.3<r/a<4$.

\section{General paradigm of the elastic interactions between colloidal particles in NLC}
   
\begin{figure}
\begin{center}
\includegraphics[width=\columnwidth]{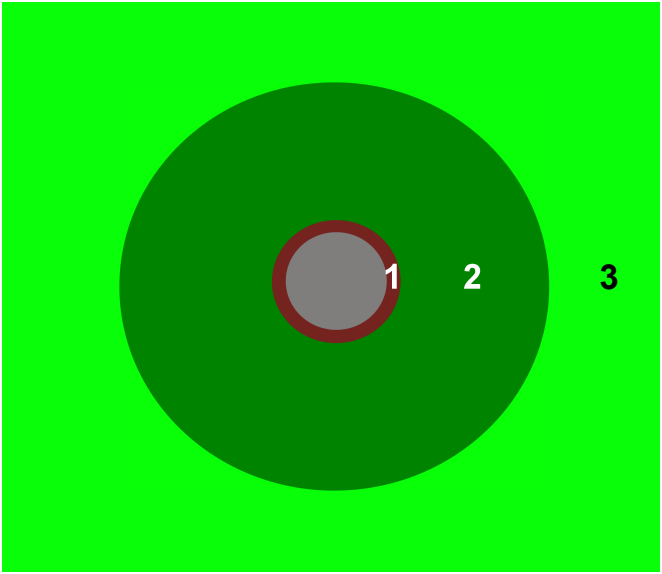}
\caption{(Color online) General structure of the nematic elastic field around the colloidal particle. Brown zone 1 - is the zone of topological defects and anharmonic deformations. The principle of superposition does not work here. Dark green zone 2 - is the zone of harmonic elastic terms of all possible orders. High order elastic terms are essential here (from 10\% to 60\% of the total deformation). Light green zone 3 - is the zone of the main elastic term. \label{zones}}
\end{center}
\end{figure}

The results obtained above help us to formulate the following picture or paradigm of the elastic interaction between colloidal particles in NLC (see Fig.\ref{zones}). 

There are three different zones around each colloidal particle. The first zone is the zone of topological defects (brown zone 1 on the Fig.\ref{zones}). Non-linear terms are very essential in this zone, the EL equation is non-linear. So that the principle of superposition does not work in the first zone. The size of the first zone is about  $0.2a-0.3a$ of the particle radius $a$, so that it is concentrated at the distances $a<r\lessapprox 1.3a$. This is in line with results obtained for the hyperbolic hedgehog and Saturn ring director configurations. In the paper \cite{lupe} it was found that hyperbolic hedgehog is located at the distance $r=1.22a$ and Saturn ring is located at the distance $r=1.08a$.

The second zone appears just after the first zone (dark green zone 2 on the Fig.\ref{zones}). In this zone anharmonic terms vanish and harmonic elastic terms of all possible from the symmetry point of view orders are born simultaneously. The director field here can be presented as multipole expansion of all possible orders and the principle of superposition is valid. All elastic terms coexist here and higher order elastic terms are essential (from 10\% to 60\% of the total deformation). It is concentrated at the approximate distance range $1.3a \lessapprox r \lessapprox 4a$. The second zone is the zone where crossover from topological defects to the main multipole moment takes place.

And the third zone is the zone of the main multipole moment, where higher order terms make contribution less than 10\% and only the first nonzero multipole moment has the dominant value. This zone extends for distances $r\gtrapprox 4a=2D$ (light green zone 3 on the Fig.\ref{zones}). The influence of the high order terms still exists in the third zone and has a contribution of approximately several percent. For instance, it makes the effective power to be non-integer like it was shown above.

\section{Conclusion}

To conclude, theoretical description of the elastic interaction between colloidal particles with the incorporation of the higher order elastic terms is proposed.

The general paradigm of the elastic interaction between colloidal particles in NLC is proposed. Each particle has three zones around itself; the first zone is the zone of topological defects where anharmonic terms are essential and the principle of superposition does not work. This zone has the size of about  $0.2a-0.3a$ of the particle radius $a$, so that it is concentrated at the distances $a<r\lessapprox 1.3a$.

The second zone appears just after the first zone. In this zone anharmonic terms quickly vanish and harmonic elastic terms of all possible from the symmetry point of view orders are born simultaneously. The director field here can be presented as multipole expansion of all possible orders and the principle of superposition is valid. The  higher order elastic terms are essential here (from 10\% to 60\% of the total deformation) and this zone is concentrated at the approximate distance range $1.3a \lessapprox r \lessapprox 4a$. It is the zone where crossover from topological defects to the main multipole moment takes place.

The last third zone is the zone of the main multipole moment, where higher order terms make a contribution of less than 10\% and only the first nonzero multipole moment has the dominant value. This zone extends for distances $r\gtrapprox 4a=2D$.

Of course all three zones exist only for particles with strong anchoring conditions at the particle surface. The first zone is absent for particles with weak anchoring and the second zone starts just from the particle's surface in this case. 

We consider the case of spherical particles with planar anchoring conditions as an example. Such particles have topological defects called boojums at the poles. We find that boojums can be described analitically via multipole expansion with accuracy up to $1/r^{7}$ and the whole spherical particle can be effectively considered as the multipole of the order 6 with multipolarity equal $2^{6}=64$. The correspondent elastic interaction between two beads with higher order elastic terms gives the angle $\theta_{min}=34.5^{\circ}$ of minimum energy between two contact beads which is close to the experimental value of $\theta_{min}=30^{\circ}$. As well high order elastic terms make the effective power of the repulsive potential to be non-integer at the range $4.5<\gamma_{eff}<5$ for different distances. The incorporation of the high order elastic terms in the confined NLC produce results that agree with experimental data as well.

The application of the higher order terms for hyperbolic hedgehog and Saturn ring director configurations is under the way. The author is grateful to Prof. B.I. Lev for fruitful discussions.

\end{document}